\documentclass[10pt,prb,twocolumn,showpacs]{revtex4}
\usepackage{graphicx}
\usepackage{epsf}
\usepackage{ifthen}
\usepackage{amsmath}
\usepackage{amssymb}
\usepackage{subfigure}
\usepackage{psfrag}
\usepackage{float}
\usepackage{subeqnarray}
\begin{document}
\title{Melting of graphene clusters}
\author{Sandeep K. Singh}
\email{SandeepKumar.Singh@ua.ac.be}
\author{M. Neek-Amal}
\email{ neekamal@srttu.edu}
\author{F.M. Peeters}
\email{Francois.Peeters@ua.ac.be}
\affiliation{Department of Physics, University of  Antwerpen, Groenenborgerlaan 171, B-2020 Antwerpen, Belgium}%

\date{\today}

\begin{abstract}
Density-functional tight-binding and classical molecular dynamics simulations are used
to investigate the structural deformations and melting of
planar carbon nano-clusters $C_{N}$ with  N=2-55. The minimum
energy configurations for different clusters are used as starting configuration for the study of the
temperature effects on the bond breaking/rotation in carbon lines
(N$<$6), carbon rings (5$<$N$<$19) and graphene nano-flakes. The larger the rings (graphene nano-flake) the higher
the transition temperature (melting point) with  ring-to-line
(perfect-to-defective) transition structures. The melting point was obtained by
using the bond energy, the Lindemann criteria, and the specific heat. We found that hydrogen-passivated
graphene nano-flakes (C$_{N}$H$_M$) have a larger melting temperature
with a much smaller dependence on its size. The edges in the graphene nano-flakes exhibit
several different meta-stable configurations (isomers) during heating before melting occurs.

\end{abstract}

\maketitle

\section{Introduction} The study of the melting of crystals is one of
the important subjects in the field of phase transitions.
Melting phenomena occurs  at the surface of bulk
materials~\cite{Dash} and needs a microscopic theory for a deep
understanding. Nano-scale molecular clusters due to their
size-dependent properties show melting processes different from
those of bulk materials and infinite size two-dimensional materials.
The melting of nano-clusters has received considerable attention recently and
it was found that nano-clusters melt typically below their
corresponding bulk melting
 temperature~\cite{Koga,Ercolessi,Lewis}. This is due to the
  higher chemical reactivity of nano-clusters which is the consequence of the increased accessible surface and the presence of more free dangling bonds.

The microscopic behavior of nano-clusters at finite temperature can
be understood theoretically  using a variety of molecular dynamics
(MD) methods~\cite{Cleveland,Rytkonen,Joshi,Chuang,Beck,Bas} and can
be  determined directly by
experiment~\cite{Schmidt,Shvartsburg,Breaux}. In most of the
simulations the microscopic structure is characterized in terms of
bond-lengths and their average fluctuations over many cycles of the
MD simulation~\cite{Rey,Rodeja,Li}.

Since the discovery of two dimensional materials, i.e.
graphene~\cite{Castro,Geim} and hexagonal boron nitride
sheet~\cite{zettle2009} the melting of these new materials have
attracted many researches~\cite{Zakhar}. The new 2D crystalline
materials respond to an increasing temperature by loosing
their lattice symmetry, e.g. Zakharchenko et al. \cite{Zakhar}
studied the high temperature behavior of graphene using atomistic
simulations. The melting temperature of graphene was estimated to be about 4900\,K. Before
melting first Stone-Wales defects  appear because of their smallest energy barrier. When increasing
temperature further eventually spaghetti-type of carbon
chains are formed that  spread in 3D.
A similar melting process can be found for carbon nanotubes using a
much smaller critical Lindemann parameter\cite{Kaiwang}. The melting
temperature of perfect single-wall carbon nanotubes (SWNTs) was
estimated to be  around 4800\,K~\cite{Kaiwang}. In graphene nano-ribbons
different types of edges (i.e. zig-zag, armchair) affect the melting process differently, e.g. \textbf{Lee et al.~\cite{Lee}
found that at 2800\,K  edge reconstruction occurs in a zig-zag
ribbon.}

In our previous work we found that the minimum energy configuration for flat carbon clusters up to N=5 atoms consists of a line  of
carbons~\cite{kosi1} (linear chain) which is in agreement with
ab-initio~\cite{raghavachari} calculations. Carbon planar rings were found for 5$<$N$<$19 and graphene nano-flakes  are minimum energy
configurations for larger N~\cite{kosi}. Here
we investigate the effect of temperature on those minimum
energy configurations and find the melting temperature of such small
flat carbon clusters, as function of the size of the clusters.

A systematic study of the size dependence of the melting temperature
is still lacking as well as the effect of H-passivation of the edge
atoms on the melting process. We will present such a study and identify the
different fundamental steps in the melting process. We found that
graphene nano-flakes have a lower transition temperature as compared
to bulk graphene and graphene nanoribbons.  \textbf{We also found that H-passivated clusters exhibit higher melting temperature than non H-passivated clusters. In all cases, } once clusters are defected they can be in different meta-stable structures (\textbf{none-planar isomers}). We will compare our results with those found for graphene and graphene nano-ribbons. \textbf{The Lindemann index increases with respect to temperature in all cases while its slope versus temperature increases  (decreases) linearly for the ring  structures (graphene nano-flakes). Furthermore, using ab-initio molecular dynamics simulation we analyse the energy change due to defect formation.}

This paper is organized as follows. In Sec. II, we introduce the
atomistic model and the simulation method. Sec. III contains our
main results and a discussion of the melting of graphene-like clusters
and H-passivated clusters. Sec. IV gives information on the
topology of the defects. The effect of defects on the total energy is introduced in Sec. V. In Sec. VI, we conclude the
paper.

\section{Simulation Method and Model}
\subsection{Minimum energy configurations}

The second-generation of Brenner reactive empirical bond order
(REBO) potential\cite{bren2} function between carbon atoms is
 used in the present work. All the parameters for the Brenner potential can be found in Ref.\cite{bren2}
 and are therefore not listed here.

\begin{figure*}
\begin{center}
\includegraphics[width=0.5\linewidth, height=12.5cm]{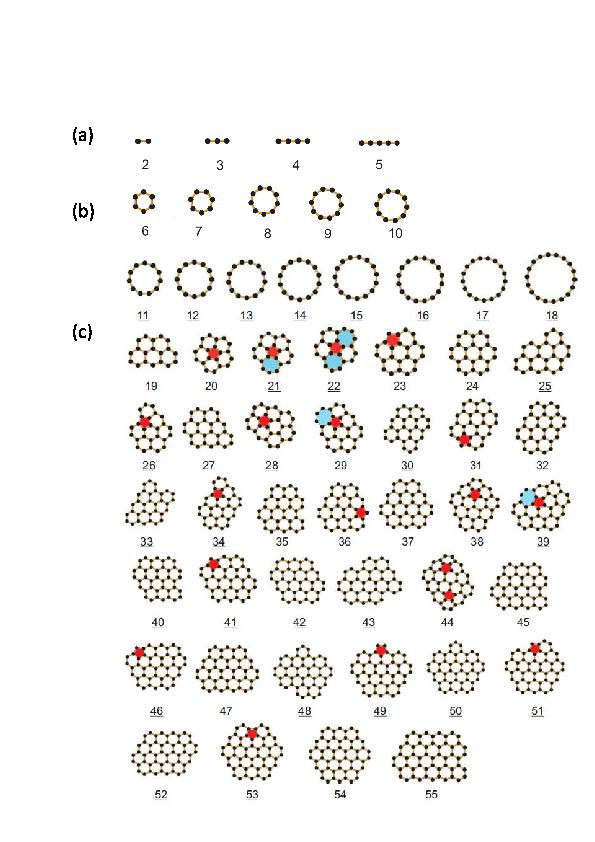}
\includegraphics[width=0.45\linewidth]{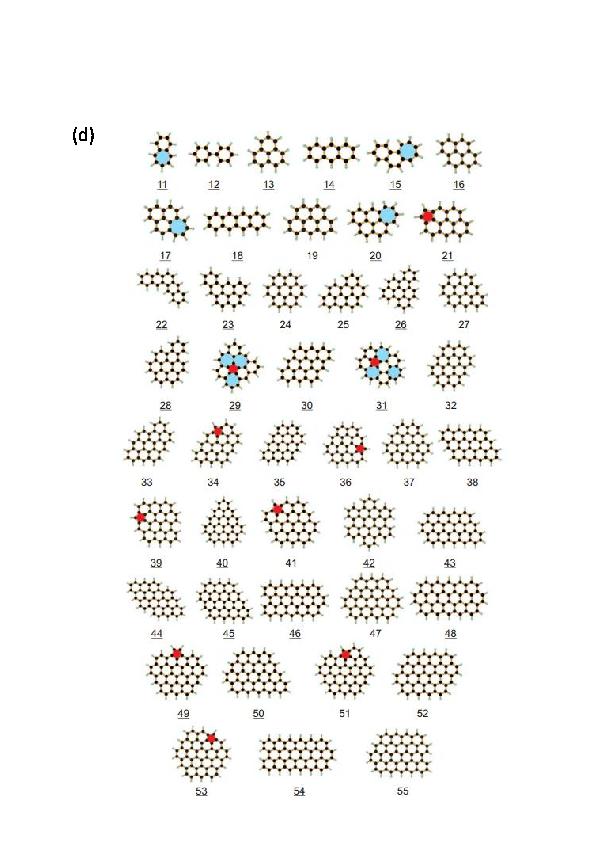}
\caption{(Color online) The investigated configurations for carbon
lines (a) carbon rings (b) and graphene nano-flakes (c). Pentagons
(heptagons) are colored red (blue). By H-passivating the structures
in (c) most of them transit to new structures (shown in (d)) which
are indicated by underlined numbers.
\label{figmodel}}
\end{center}
\end{figure*}

In Fig.~\ref{figmodel} we depict the minimum energy configurations
for carbon clusters which are carbon lines up to 5 atoms (Fig.~\ref{figmodel}(a)), carbon rings for up to 18 atoms
(Fig.~\ref{figmodel}(b)), graphene nano-flakes up to 55 atoms
(Fig.~\ref{figmodel}(c)) and hydrogen passivated graphene
nano-flakes (Fig.~\ref{figmodel}(d)). These configurations were
obtained using conjugate gradient minimization method in our
previous works~\cite{kosi,kosi1}. The carbon line structures are
energetically favorable structures among other possible geometries (isomers)
which are shown in Fig. 1 of Ref.~\cite{kosi1}, i.e. two for $C_3$, 6
for $C_4$ and 11 for $C_5$. Among all possible carbon nano-clusters (isomers) for 5$<$N$<$19
atoms the ground state are a single ring, see
Fig.~\ref{figmodel}(b). Increasing the number of carbon atoms, graphene nano-flakes are formed which can have pentagon and heptagon defects in addition to common hexagons, see Fig.~\ref{figmodel}(c). Notice that by passivating the dangling bonds by hydrogens in graphene flakes some structural deformations are possible. In Fig.~\ref{figmodel}(d) the minimum energy configurations for hydrogen passivated graphene nano-flakes (which were obtained by passivating the structures in Fig.~\ref{figmodel}(c)) are shown. \textbf{It is interesting to note that most of these minimum configuration structures have zig-zag edges which is due to the higher stability of these kind of edges as compared to arm-chair edges~\cite{Seifert}.}

In the present work we study the temperature effects on the structural
transition and melting properties of these minimum energy
configurations. Using molecular dynamic simulations, we obtain the
new configuration of the above mentioned clusters at a given
temperature T. This temperature is maintained during the whole
simulation by the Langevin thermostat~\cite{rui}. The  MD time step
was taken to be 0.5 fs. Different properties of the cluster were
measured during the MD simulation of $10^{6}$ MD steps (500 ps) at
fixed temperature.

\subsection{Density-functional tight-binding molecular dynamics}
 In order to have an independent test of the results obtained from the bond
order potential  for the melting of graphene nano-flakes, we also performed
independent calculations using the DFTB/MD (density-functional based
tight-binding molecular dynamics) approach which is a QM/MD
technique based on a tight binding method
 using an approximate density-functional formalism~\cite{Porezag, Elstner, Frauenheim}.
 DFTB passed several benchmark tests with first principle density functional theory (DFT)~\cite{Porezag, Zheng, Alberto} for
  carbon structures.  Alberto et al.~\cite{Alberto} showed that DFTB accurately reproduced the structures and energies for a range of point defects such as
 vacancies and Stone-Wales defects in graphene. Migration barriers for vacancies and Stone-Wales defect formation barriers are also
 accurately reproduced. \textbf{Kuc et al~\cite{Seifert} studied the stability of graphene nano-flakes using DFTB and by comparing their
 results with DFT,  good agreement was found between the two methods.} Although this method is two orders of magnitude
 faster than DFT but for the purpose of this work where we will study
 about 90 different configurations it will be computationally
expensive. Therefore, we will use DFTB for a few N-values in order to show the accuracy of our classical MD simulation by considering one of the line carbons,
 two of the rings, five of graphene nano-flakes and six of H-passivated
 systems.

\subsection{ Lindemann criterion and specific heat}
The root-mean-square relative bond length variance (Lindemann
criterion) in addition to the caloric curve gives a reasonable
computational method for determining the melting point of nano-clusters. It is sensitive to
any change in the bond lengths at the microscopic scale. The Lindemann
criterion~\cite{March, Ziman} is often used in molecular dynamics
and Monte Carlo simulations in order to estimate the melting temperature in three dimensional bulk
systems~\cite{Blome, Shen}, two dimensional
materials~\cite{Zakhar} and nano-clusters~\cite{erik}.  We used the
distance-fluctuation of the Lindemann index ($\delta$) in order to
identify  the melting temperature of our nano-clusters.
   For a system of N atoms, the local Lindemann index for
    the ${\imath}^{th}$ atom in the system is defined as~\cite{Zhou,Ding}
\begin{equation}
\delta_{i}=\frac{1}{N-1}\sum_{j(\neq i)}\frac{\sqrt{\langle r_{ij}^{2} \rangle_{T}-\langle r_{ij}\rangle_{T}^{2}}}{\langle r_{ij}\rangle_{T}}
\end{equation}
and the system-average Lindemann index is then given by
\begin{equation}
\delta=\frac{1}{N}\sum_{i} \delta_{i}
\end{equation}
where $r_{ij}$ is the distance between the ${\imath}^{th}$ and ${\jmath}^{th}$ atoms, N is the number of atoms
and $\langle \cdot \cdot \cdot\rangle_{T}$ denotes the thermal average at temperature T. The Lindemann index
 ~\cite{Lindemann} depends on the specific system and its size which varies in the range
 0.03-0.15, e.g.~ it was recently~\cite{Zhou} applied to nanoparticles and homopolymers
 and found to be in the range of 0.03-0.05, for Ni nanoclusters it was found
 to be around 0.08~\cite{erik} and for carbon nanotubes about 0.03~\cite{Kaiwang}.

For sufficiently low temperature there is no structural
transition and the atoms exhibit thermal fluctuation  around the $T=0$ equilibrium
position. The oscillation amplitude increases linearly with
temperature due to Hooke's regime for the atomic vibrations leading
to a linear increase of the Lindemann index with T.  At higher
temperature, the \textbf{anharmonic vibrations} (non-linear effects) become important and the
Lindemann index exhibits a nonlinear dependence on T. The particle oscillation amplitude
increases faster than linear with T, but the system does not melt
yet, since the arrangement of atoms have still some ordered structure,
i.e. solid-liquid coexistence state. For small nano-clusters the
latter is related to not well defined small three dimensional
structures.  In general, melting occurs when the Lindemann index increases
 very sharply with T over a small T-range. In this study we will
 assume that the melting point is around the sharp jump in  $\delta$, i.e. when the system becomes almost a random coil.
We will show that the Lindemann index \textbf{adequately indicates the structural}
deformation (melting-like transition) of carbon
 nano-clusters. The obtained linear regime in $\delta$ is smoother
 than some of the previous studies~\cite{erik} for small  nano-clusters.
   Therefore, we will not only use the critical value of $\delta$
to  determine the melting point  but we will also pay particular
attention to the temperature dependence of the Lindemann index when
identifying the melting temperature.

In addition to the Lindmann index, the total energy (caloric curve) and
specific heat variation versus T are two common quantities which can
be used to determine the phase transition. We calculated the specific
heat $C_{P}$ using the equation\cite{Hsu}
\begin{equation}
C_{P}(T)=\frac{\langle E^{2}_{total} \rangle_{T}-\langle E_{total}\rangle_{T}^{2}}{k_{B} T^{2}},
\end{equation}
where $E_{total}=\sum_{i}\frac{1}{2} m_{i} v_{i}^{2} + E_{P}$.
The average potential energy of the
 system was calculated as a function of temperature. In the crystalline state the total energy of the
 system increases almost linearly with temperature, and then after the critical temperature is reached,
 it increases more steeply which is a signature of melting. We will
 show that for graphene nano-flake with 54 carbon atoms (C$_{54}$ and C$_{54}$H$_{20}$) energy and heat capacity
 calculations are found to be consistent with the results for the melting temperature that we obtained  from the analysis
 of the Lindemann index.

\section{Results and Discussion}
\subsection{ Energy }

\begin{figure}[htb]
\centering
    \includegraphics[width=8cm]{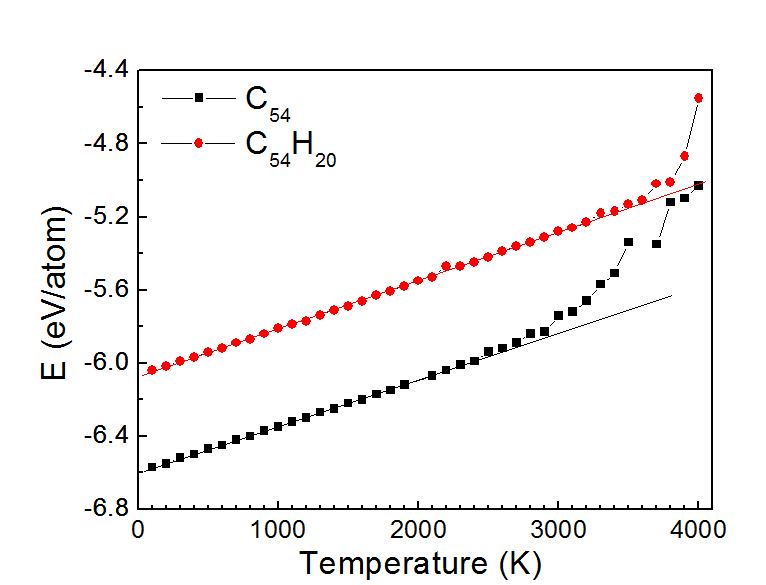}
  \caption{(Color online)
The temperature dependence of the total energy of the graphene
nanoflake  C$_{54}$  and  the H-passivated C$_{54}$H$_{20}$ using the REBO potential.
} \label{54_pot}
\end{figure}

\begin{figure*}[htb]
\centering
    \includegraphics[width=8cm,height=6.4cm]{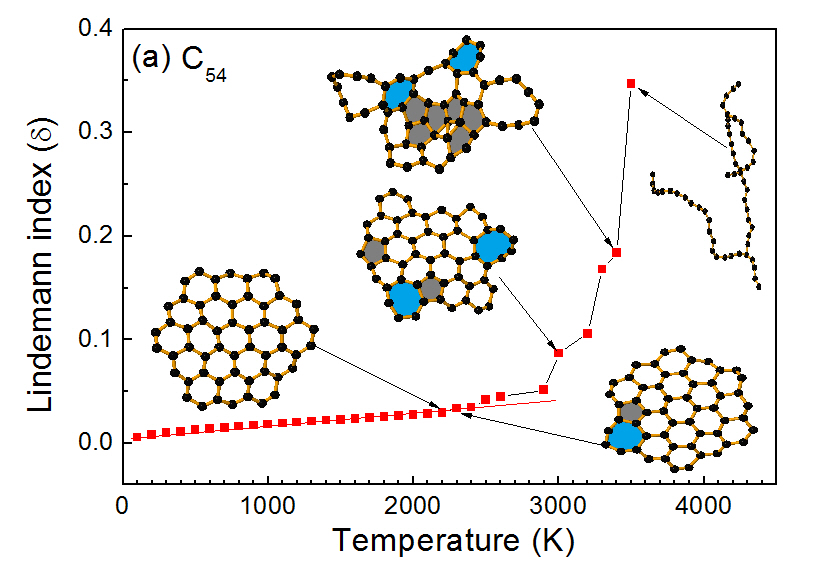}
    \includegraphics[width=8cm]{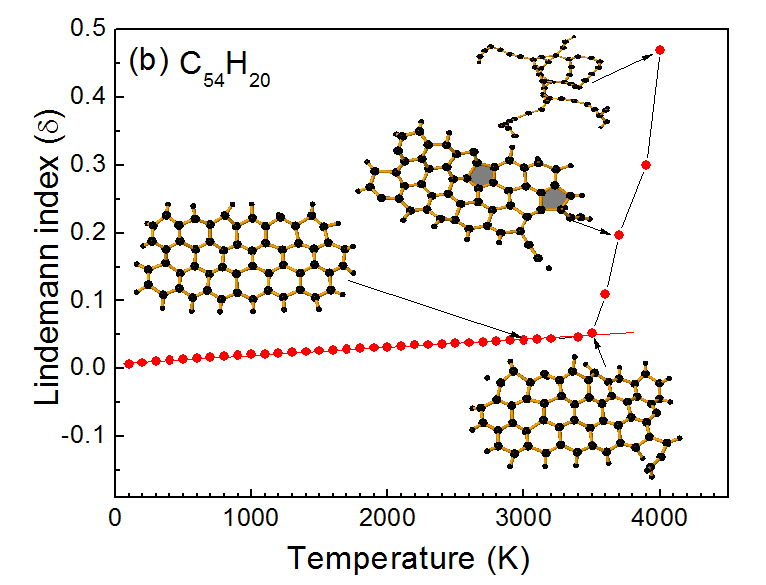}
  \caption{(Color online)
The temperature dependence of the Lindemann index for the cluster
(a) without H-passivated C$_{54}$ and  (b) with H-passivated  C$_{54}$H$_{20}$. The insets show typical C (a) and C-H (b) atoms configurations where the solid
areas indicate topological defects. }\label{54_delta}
\end{figure*}

\begin{figure*}[htb]
\centering
    \includegraphics[width=8cm]{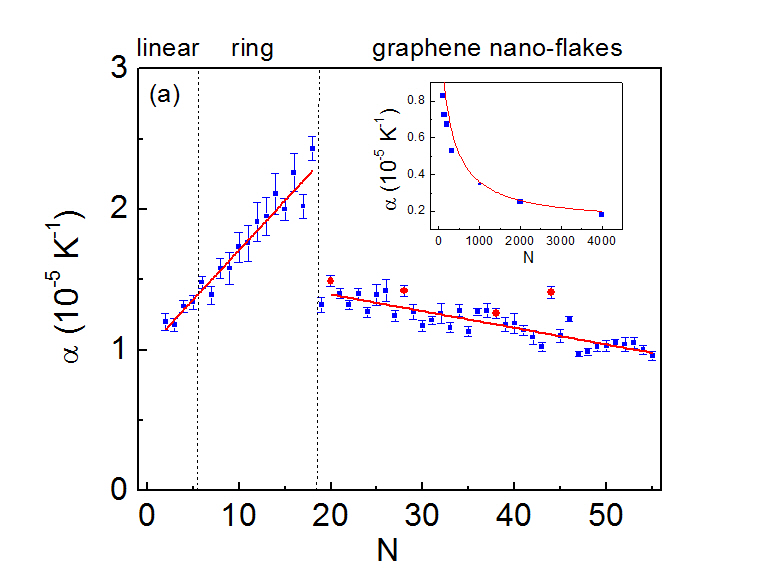}
    \includegraphics[width=8cm]{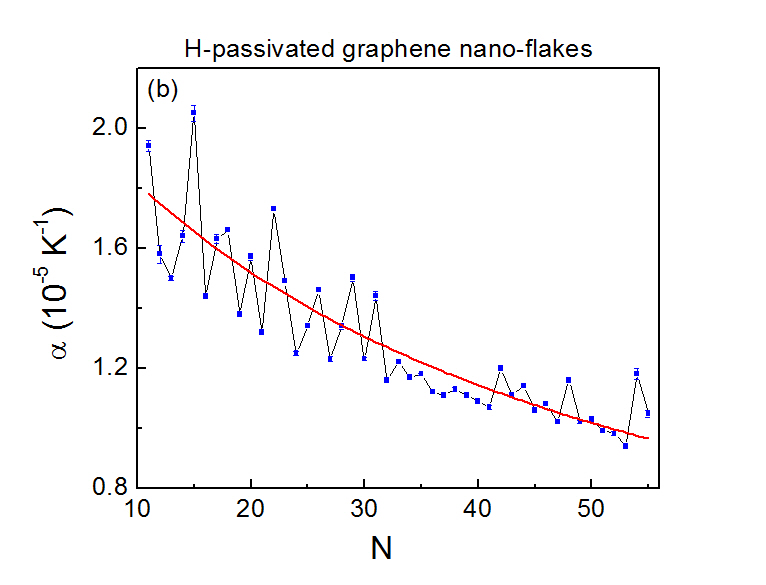}
  \caption{(Color online)
The low temperature rate of the Lindmann index versus the number of atoms in (a) graphene nano-flakes and (b) H-passivated graphene nano-flakes.
The red solid lines are linear fits to the average behavior of $\alpha$.
The bowl-like clusters are shown by red solid circles (a). In the inset, the same quantity
 versus the number of atoms in graphene nano-flakes for N up to 4000 (a).
}\label{alpha}
\end{figure*}

\begin{figure}[htb]
\centering
    \includegraphics[width=8cm]{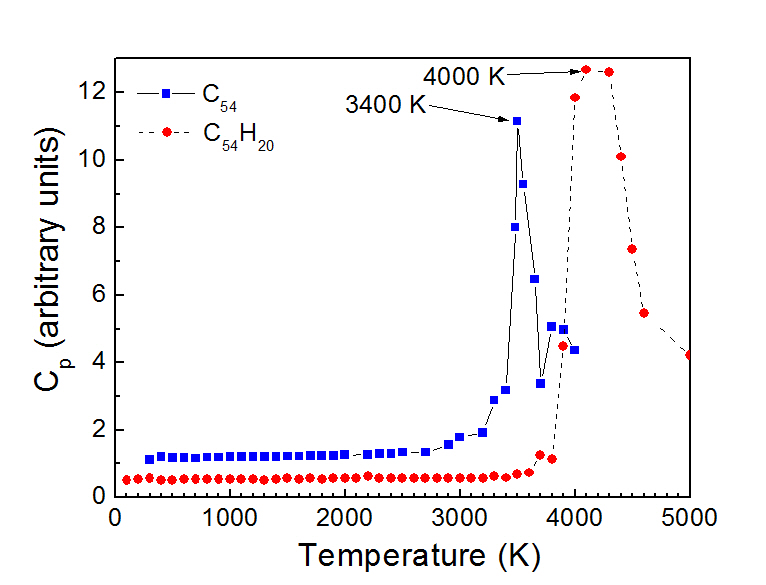}
  \caption{(Color online)
The temperature dependence of the specific heat for C$_{54}$ and
C$_{54}$H$_{20}$. }\label{54_spe}
\end{figure}

The temperature dependence of the total energy of   graphene
nano-flakes (C$_{54}$ and C$_{54}$H$_{20}$) are depicted
in Fig.~\ref{54_pot} using the Brenner potential. For C$_{54}$ the
energy increases linearly at low temperatures
and starts to deviate from the linear behavior around T=2300\,K due to
the reconstruction of the zigzag edges and the formation of
pentagon-heptagon (5-7) defects. It indicates that the first
nucleation of melting starts around 2300\,K and modifies the
edges. As temperature increases, the formation of pentagon,
heptagon, 5-7 defects or 5-8-5 are possible and eventually large
ring structures results in  a dramatic increase of the  energy. Above T=~3400 K, there is a sharp increase  in the
 energy showing a completely molten structure.

For C$_{54}$H$_{20}$, passivation removes the dangling orbitals
of the  C atoms at the edge, lowering the reactivity, and
increasing the stability of the cluster.  It was found that
unlike previous case (without H-passivated clusters), the
H-passivated clusters keep their initial atomic arrangement up to
higher temperature and therefore no noticeable change in the
geometry was found for temperatures up to T=3200 $K$ for C$_{54}$H$_{20}$. The binding
energy of the H-passivated clusters are larger than for
 non-passivated clusters. As temperature increases further to T=3500\,K,  some hydrogen atoms start to dissociate and
 finally the clusters convert into hydrocarbon chains around T=4000\,K showing larger melting
 temperature than the corresponding non-passivated cluster which has a melting temperature of 3400\,K (see Fig.~\ref{54_pot}).

\subsection{ Lindemann index}

Fig.~\ref{54_delta}  displays  the variation of the Lindemann index with
temperature for (a) C$_{54}$ and (b) C$_{54}$H$_{20}$. The corresponding
structures during heating for a few typical temperatures are shown in
the insets.  The slope of the function $\delta(T)$ (i.e.
$\alpha=\frac{d\delta}{dT}$) is plotted in Fig.~\ref{alpha} for all
studied structures of Fig.~\ref{figmodel}. The value of $\alpha$
(before reaching the melting point) increases monotonically
 for the carbon lines (Fig.~\ref{figmodel}(a)) and carbon rings (Fig.~\ref{figmodel}(b)).
\textbf{This is an indication of keeping the initial configuration while non-linear effects indicate defect formation, e.g.  5-7 defects in C$_{54}$.}
The increase of $\alpha$ is fitted
in Fig.~\ref{alpha} by the red line for $N \leq 18$ which is given  by the
function $\alpha(N)=a+b \times N$, where: a=1.005 ($\pm0.032) \times 10^{-5} K^{-1}$ and
b=0.070 ($\pm 0.004) \times 10^{-5} K^{-1}$. Note that
increasing temperature forces the system to be out-of-planed, e.g.
the rings at finite temperature are not circles and become deformed
ellipsoids in 3D.

For N $\geq$ 19, there is a sudden decrease
in  $\alpha$ of size $\Delta \alpha=1.11 \times 10^{-5} K^{-1}$ due to
the  strong $sp^2$ bonds within the graphene like clusters (instead of
simple covalent bonds in the carbon lines and rings) and a decreasing number of dangling bonds.
 The average behavior is
fitted by the red line $\alpha(N)=a+b \times N$, where: a= 1.63($\pm0.06) \times 10^{-5} K^{-1}$ and b=-0.012 ($\pm0.001)  \times 10^{-5} K^{-1}$.  For
bowl like clusters (N=20, 28, 38, 44), due to the presence of topological  defective pentagon
inside the cluster, the $\alpha$ value is larger as compared to their neighbor
clusters. Therefore the important message is that dangling bonds and any
  kind of defects enhance  anharmonic effects.

In order to investigate the effect of large size samples we also
calculated $\alpha$ for a few large graphene nano-flakes
and found that $\alpha$ decreases with N (see inset in
Fig.~\ref{alpha}). The  maximum considered size of graphene nano-flakes had
4000 atoms.  \textbf{For large N, one expects saturation of $\alpha$, thus a line with negative slope which we fitted for
 19 $\leq$ N $\leq$ 55 should not be applicable.} Therefore, we used the fit $\alpha(N)=\alpha(\infty)+a/(1 + b\times N)$, where: $\alpha(\infty)= 0.134 \times 10^{-5} K^{-1}$, a=1.213 ($\pm0.005) \times 10^{-5} K^{-1}$
and b=48.7 ($\pm 4.9) \times 10^{-4}$ on large clusters. These results clearly indicate  that $\delta$ for  small
graphene nano-flakes is considerably larger than for larger flakes and
graphene. Although the  Lindemann index was defined initially
in the thermodynamical limit (bulk material) we show that
it is also a good parameter to investigate the effect of temperature
and melting of nano size systems.

For completeness, we calculated $\alpha$ for H-passivated  graphene nano-flakes
for 11$\leq$ N $\leq$ 55 number of C-atoms and found that $\alpha$, on average,  decreases with N (see Fig.~\ref{alpha}(b)). The average behavior is
fitted by the red line $\alpha(N)=a/(1 + b\times N)$, where: a=2.262 ($\pm0.115) \times 10^{-5} K^{-1}$
and b=0.024 ($\pm 0.003) \times 10^{-5}$. The rapid decrease in the latter fit (Fig.~\ref{alpha}(b)) is an indication of the role of H-passivation in making the graphene nano-flakes more stable against temperature for larger N.

\subsection{Specific heat}
The calculated specific heat curve for C$_{54}$ is shown in
Fig~\ref{54_spe}. A clear peak is observed in the specific heat
with a maximum around 3400\,K which we identify as the melting
temperature, and which is close to the results from the analysis
using the Lindemann index. The specific heat curve is also
 shown for C$_{54}$H$_{20}$ (see red symbols in Fig.~\ref{54_spe}) and
 displays a peak around T=4000\,K which is identified as the melting temperature, showing good agreement
 with the result of the previous caloric curve (see Fig.~\ref{54_pot}). \textbf{A discontinuity or a sharp peak \
 in the heat capacity is a clear indication of a phase transition. However, here
 this is not exactly a  solid-to-liquid like transition, but rather a  nano-flake to random-coil transition. }

\begin{figure*}[htb]
\centering
    \includegraphics[width=8cm,height=6.4cm]{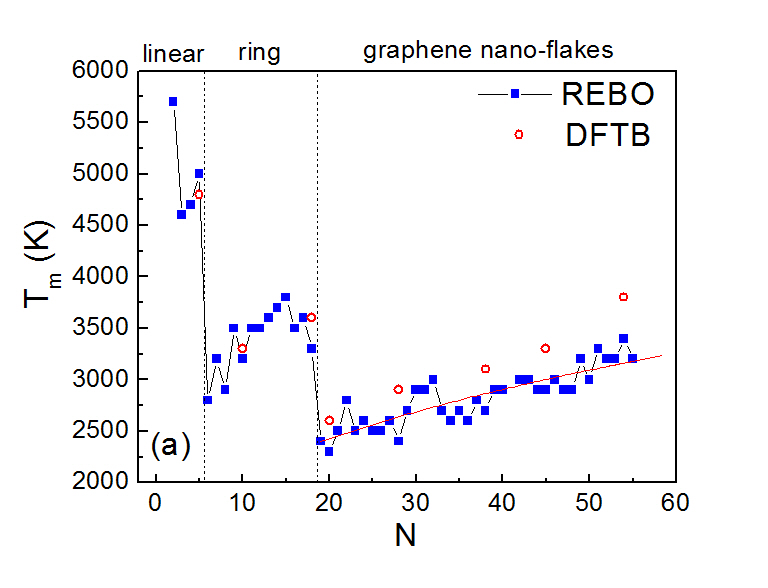}
    \includegraphics[width=8cm]{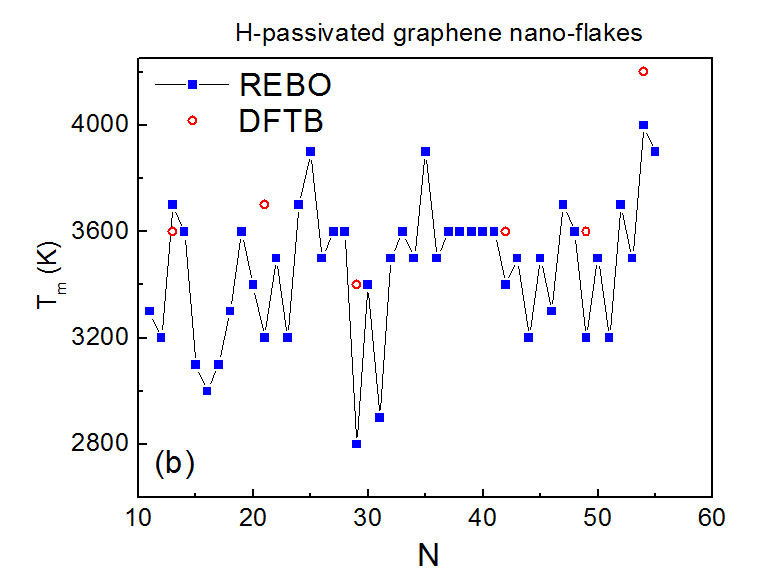}
  \caption{ (Color online)
The melting temperature versus  the number of atoms using Brenner (square symbols) and DFTB (open circles) for (a) nano-graphene and (b) hydrogenated nano-graphene. The error
bar is 50\,K.}\label{melting}
\end{figure*}

\begin{table}[tp]%
\caption{Melting temperature for large clusters}
\begin{tabular}{l  l}
\hline
$N$                & $T_{m} (K)$                  \\
 \hline
98                 & 3800 \\
142                & 4000  \\
194                & 4100     \\
322                & 4200     \\
1000                & 4400     \\
graphene           & 5500           \\
 \hline
\end{tabular}
\end{table}

\subsection{Melting-temperature}
The melting temperature for all studied clusters $C_{N} (2\leq N
\leq 55)$ is shown in Fig.~\ref{melting}(a). The different parts
separated by vertical dashed lines correspond to the three set of
systems shown in Figs.~\ref{figmodel}(a-c). For small linear structures
except for the carbon dimer the
 melting temperature increases with the number of atoms (linear chains have bond
breaking).

In order to check if for large N we approach the melting temperature of bulk
graphene, we also calculated the melting temperature of three large
graphene nano-flakes and graphene (by performing a simulation having
4000 atoms with periodic boundary condition with NPT ensemble) and
presented the results in Table I. From Table I it is clear that
large flakes approach slowly the melting temperature of graphene, i.e.
5500\, K. As a comparison the melting temperature reported in
Ref.~\cite{Zakhar}  by using the LCBOPII potential was 4900 K and in Ref.~\cite{Brad} using REBO was 5200 K.

\textbf{We fitted the melting temperature for graphene like clusters
(red curve in Fig.~\ref{melting}(a)) by the function
$T_{m}(N)=T_{m}^{bulk}-(a/(b+N))$ where $a=417(\pm47) \times 10^{3}
K$, $b=119.34(\pm17.71)$ and $T_{m}^{bulk}=5500 K$ for graphene was
taken from our simulation. We included the results of Table I in
this fit. We also calculated the melting temperature using DFTB for the clusters
with N=5, 10, 18, 20, 28, 38, 45, and 54 (C$_{54}$) which are represented by the
open red circles in Fig.~\ref{melting}(a). They are found to slightly
overestimate the melting temperature but exhibit clearly the same N-dependent trend.}

For completeness, we also calculated the melting temperature of
H-passivated clusters for N=11-55 C-atoms (Fig.~\ref{melting}(b)).
The melting temperature fluctuates around
T=3500\,K (note that on average T$_{m}$ increases slowly with N with
large fluctuations imposed on it) for most of the clusters with minima for N=29 and N=31
C-atoms due to the large number of defects in their structures. The
clusters which have pentagon defects on the boundary usually have
lower melting temperature then the others. Here, the melting temperature
for the clusters N=13, 21, 29, 42, 49 and 54 were also calculated using
DFTB (open red circles in  Fig.~\ref{melting}(b)).
In most cases the DFTB results are close to the Brenner potential
results indicating that the Brenner bond order potential is a useful
specialized potential for thermal effects in hydrocarbons.

\section{Topology of Defects}
\begin{figure*}[htb]
\centering
   \includegraphics[width=13cm,height=18cm]{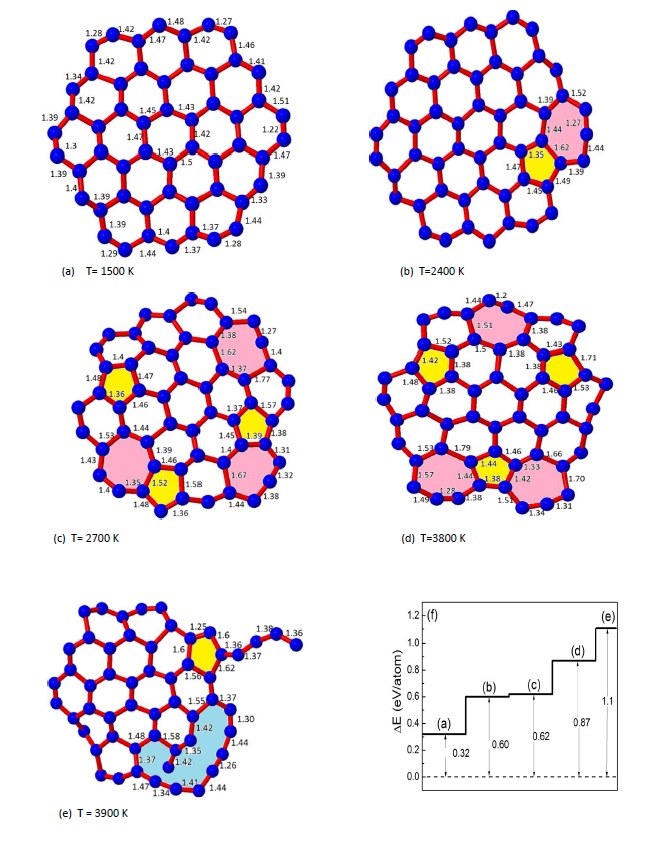}
  \caption{ (Color online)
Five different snap shots of the cluster  C$_{54}$ at different temperature (a-e). Bond
lengths are indicated in the figure and the colored areas indicate defect structures.
(f) The corresponding energy diagram for the five snap
shots.}\label{5-7}
\end{figure*}

\begin{figure*}[htb]
\centering
  \includegraphics[width=14cm,height=18cm]{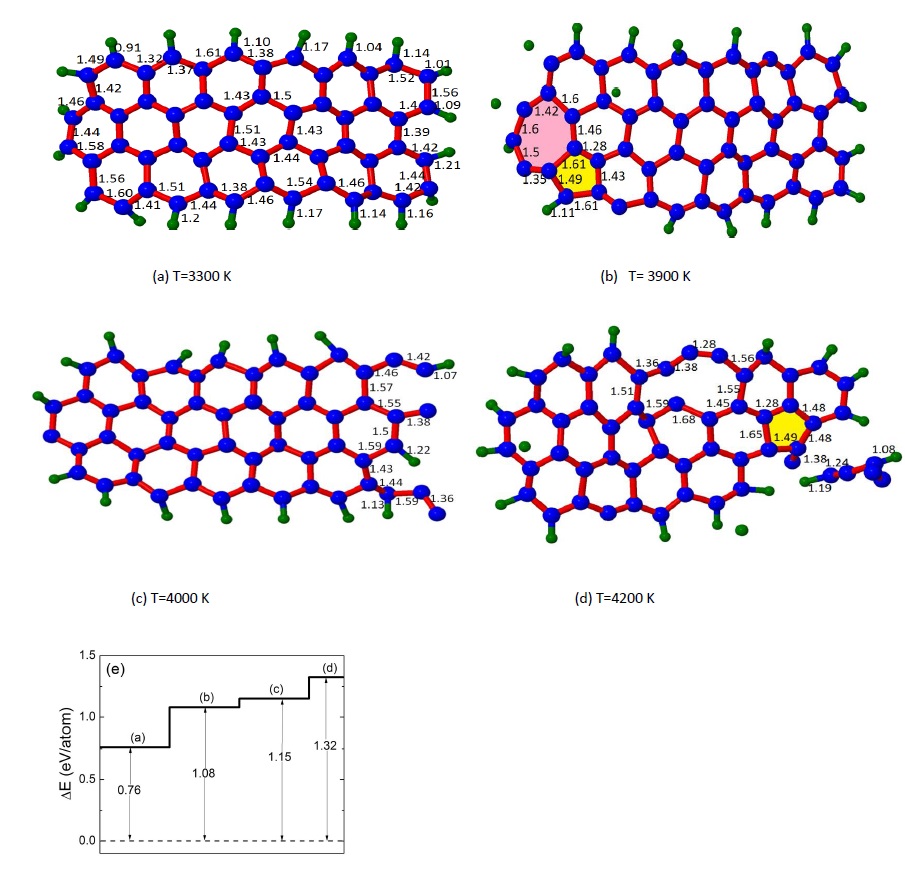}
  \caption{ (Color online)
Four different snap shots of the cluster  C$_{54}$H$_{20}$ at different temperature (a-d). Bond
lengths are indicated in the figure and the colored areas indicate defect structures.
(e) The corresponding energy diagram for the four snap
shots.}
\label{5-7h}
\end{figure*}

In this section we consider the topology of some of the defects which
are created during the melting process of C$_{54}$ using DFTB
calculations. We found that these defects have a pronounced effect
on the melting behavior of the system whose mechanism is
different from both graphene and graphene nanoribbons. At low
temperature the probability of defect creation is small and the
flakes remains perfect which for this low temperature range is
similar to graphene~\cite{Zakhar} and graphene
nanoribbon~\cite{Koskinen}. Increasing temperature (above
about 2300\,K) the energy increases and the system can overpass
certain potential barriers and we found that during the molecular
dynamics simulation the system transits from one
metastable state to another, see Figs.~\ref{5-7}(a-b). We show five
snap shots of C$_{54}$ at different temperature in Fig.~\ref{5-7}.
The transition temperature at which the edge reconstruction occurs, i.e. T$\sim$ 2400\,K is lower than those
found for defect formation in  graphene, i.e.~3800\,K and edge reconstruction in graphene
nano-ribbons~\cite{Lee}, i.e. 2800\,K. As seen in
Fig.~\ref{5-7}(b) the first defected structure has one heptagon and
one pentagon at the edge with different bond lengths. The energy
difference between non-defected (Fig.~\ref{5-7}(a)) and defected
(Fig.~\ref{5-7}(b)) one is about 0.28~\,eV/atom which is due to
the 900~\,K difference in the temperature between these two structures.
In Fig.~\ref{5-7} the next defected configuration due to a 1200\,K
increase in temperature is shown which has more heptagon and
pentagon defects (colored parts). By increasing temperature further we found
some transition in the defected parts and even an edge reconstruction
from heptagon to hexagon and vise versa until the appearance of
a tail-like part in the cluster, Fig.~\ref{5-7}(e). The melting
temperature for C$_{54}$ as shown in Fig.~\ref{melting}(a)
(open circle symbol) is around 3900\,K. This melting temperature is
lower than those found for graphene. The larger the graphene
nano-flake the higher the melting temperature. For small graphene
nano-flakes the larger number of dangling bonds results in a  lower
melting temperature and larger boundary effects. The energy diagram
for five snap shots is depicted in Fig.~\ref{5-7}(f). The presented
energy is the difference between the total energy of the system at
given temperature and the zero temperature total energy for C$_{54}$.

In Figs.~\ref{5-7h}(a-d) we show the temperature effect on
C$_{54}$H$_{20}$. In Fig.~\ref{5-7h}(e) the corresponding energy
diagram is shown. Hydrogens become released at T=3900\,K. Notice that
DFTB calculations for the shown configurations in
Figs.~\ref{5-7h}(a-d) were performed separately, i.e. we do not
increase temperature of sample Fig.~\ref{5-7h}(a) in order to obtain
Fig.~\ref{5-7h}(b), but instead we performed four different
calculations for these four snap shots. Thus we do not expect that
we have sequential configurations.

\section{Conclusion} Using  molecular dynamics simulation and the
Lindemann index for melting supplemented with results for the total energy and the
specific heat, we investigated the melting of carbon nano clusters. The
melting temperature of small carbon flakes is lower than those
for graphene and graphene nano-ribbons. The Lindemann index is
sensitive to temperature and is a good quantity
 for determining when structural deformations of the clusters start to occur.  The melting
temperature of small flakes on average increases versus the number of atoms in
carbon nano clusters. All clusters investigated show premelting
behavior with different premelting intervals. For certain N-values
defects are already present inside the cluster which lowers the
melting temperature. H-passivated clusters have a higher melting
temperature than the non H-passivated clusters with the same number
of C-atoms. \textbf{The melting temperature for H-passivated clusters
is larger than for non-passivated clusters}. Our simulation
results also help to understand the formation of defects (\textbf{due to the increase of temperature}) in the graphene
nano-flake which can then be applied to understand the growth and
thermal treatment of nanographene. We supplemented our analysis by DFTB calculations
which confirm the N-dependence of the melting temperature.

\section{ACKNOWLEDGMENTS} This work was
supported by the EU-Marie Curie IIF postdoc Fellowship/299855
(for M.N.-A.), the ESF-EuroGRAPHENE project CONGRAN, the Flemish
Science Foundation (FWO-Vl), and the Methusalem Foundation of the Flemish Government.

\end{document}